\documentclass[aps,pre,twocolumn,amsmath,amssymb,floatfix,eqsecnum,showpacs]{revtex4}

\usepackage{graphicx}
\usepackage{dcolumn}
\usepackage{tabularx}

\begin{document}

\title{Current Redistribution in Resistor Networks: Fat-Tail Statistics
in Regular and Small-World Networks}

\author{J\"org Lehmann}
\email{joerg.lehmann@ch.abb.com}
\affiliation{ABB Switzerland Ltd., Corporate Research, Segelhofstrasse 1K, CH-5405 Baden-D\"attwil, Switzerland}
\author{Jakob Bernasconi}
\affiliation{ABB Switzerland Ltd., Corporate Research, Segelhofstrasse 1K, CH-5405 Baden-D\"attwil, Switzerland}

\date{\today}

\begin{abstract} The redistribution of electrical currents in resistor
networks after single-bond failures is analyzed in terms of
current-redistribution factors that are shown to depend only on the
topology of the network and on the values of the bond resistances. We
investigate the properties of these current-redistribution factors for
regular network topologies (e.g. $d$-dimensional hypercubic lattices) as
well as for small-world networks. In particular, we find that the
statistics of the current redistribution factors exhibits a fat-tail
behavior, which reflects the long-range nature of the current redistribution
as determined by Kirchhoff's circuit laws. \end{abstract}


\maketitle

\section{Introduction}

The stability assessment of complex, interconnected systems such as
technical infrastructure networks for data, traffic or electrical power
requires the understanding of the failure-spreading mechanisms in such
systems~\cite{Newman2011,Strogatz2001,Barthelemy2011}. In one common scenario, failures propagate via the successive
overloading of the elements forming the system: An initial damage leads
to an increased stress on the remaining, undamaged parts of the system,
thereby inducing subsequent failures. Ultimately, such a failure cascade
might lead to a total breakdown~\cite{Dobson2007}. In such a scenario, one of the most
important quantities influencing the system stability is the
post-failure load redistribution, i.e., how the failing load influences
the other elements. This load-redistribution, of course, depends on the
specific system. In the present work, we shall consider a model
where the loads are given by currents across bonds in
a resistor network so that the load redistribution after removing one
bond is determined by Kirchoff's circuit laws. It will turn out that
the current redistribution in such networks can be described by a
linear rule of the form
\begin{equation}
  \label{eq:LRD}
  I_{ij}' = I_{ij} + I_{mn} \, \Delta_{ij,mn}\,,
\end{equation}
where $I'_{ij}$ and $I_{ij}$, respectively, refer to the current
across the bond~$i$--$j$ after and before the failure of
bond~$m$--$n$ with current~$I_{mn}$. The
\textit{current-redistribution factors}~$\Delta_{ij,mn}$ reflect the
influence the failing bond has on the currents flowing through the
remaining bonds. We will show that the $\Delta_{ij,mn}$ are
independent of the pre-failure currents and thus only depend on the
topology of the network and the value of the bond resistances.

Our motivation behind the study of resistor networks is manifold:
Firstly, within the so-called DC power-flow approximation, which is
commonly used also in practical applications, the calculation of the
flows in an AC power transmission grid can be mapped to the
calculation of currents in a resistor network~\cite{Wood1996}.  The
stability of power grids with respect to a cascading overloading of
its transmission lines is thus linked to that of the corresponding
resistor network~\footnote{We remark that in the context of power
  transmission grids, the current-redistribution
  factors~$\Delta_{ij,mn}$ are commonly referred to as line-outage
  distribution factors.}.

Secondly, an arbitrary weighted graph can be interpreted as a resistor
network---with resistances equal to the weights. An analysis of the
current redistribution and, in particular, of the redistribution
factors~$\Delta_{ij,mn}$ in resistor networks can thus give us insight
into the structure and topology of the corresponding weighted graphs.
A commonly investigated quantity in such graphs is the so-called
two-point resistance, which turns out to correspond directly to a
distance measure between arbitrary nodes of the graph, the so-called
resistance distance~\cite{Klein1993}. For graphs with an underlying
``geometrical'' structure, the resistance distance has been analyzed,
e.g., by Cserti~\cite{Cserti2000}. By using lattice Green functions,
Cserti calculates the two-point resistance $R_{ij}$ between arbitrary
nodes $i$ and $j$ in infinite $d$-dimensional hypercubic, rectangular,
triangular and honeycomb lattices. In particular, the dependence of
$R_{ij}$ on the spatial distance between nodes is analyzed in detail
for a square lattice, and the asymptotic form of $R_{ij}$ for large
separations between $i$ and $j$ is determined. Korniss et
al.~\cite{Korniss2006}, on the other hand, study the behavior of
$R_{ij}$ in small-world resistor networks and derive an expression for
the asymptotic behavior of the disorder-averaged two-point resistance.

In some sense complementary to the two-point resistance and other
node-centric measures, the redistribution factors~$\Delta_{ij,mn}$
provide information about the influence between two arbitrary
edges~\cite{Schaub2014}. In the present paper, we shall consider
regular as well as small-world resistor networks and investigate how
the current redistribution factors~$\Delta_{ij,mn}$ depend on the
distance $d$ between the edges $i$--$j$ and $m$--$n$.

An equally important aspect of our present investigations is the
statistics of the current-redistribution factors~$\Delta_{ij,mn}$ in
different network topologies. Our corresponding interest is motivated
by our recent study~\cite{Lehmann2010}, where we have introduced and
analyzed a new class of stochastic load-redistribution models for
cascading failure propagation. In these models, the load of a failing
element is redistributed according to a load-redistribution rule of
the form of Eq.~\eqref{eq:LRD} but interpreted in a stochastic sense,
i.e., the current-redistribution factors are assumed to be random
variables, drawn independently from a given distribution
$\varrho(\Delta)$. In the analysis of Ref.~\cite{Lehmann2010}, we have
considered a very simple, generic model for a power grid by assuming a
bimodal distribution of the $\Delta$-values. One of the main aspects
of the present work is thus to obtain a better understanding of the
statistics of the current redistribution in different network
topologies.

That such a stochastic model can indeed describe important stability
properties of a system has been shown in Ref.~\cite{Lehmann2010a}.
There, the fracture of materials has been analyzed by means of a fiber
bundle model with a stochastic stress redistribution of the form of
Eq.~\eqref{eq:LRD} that had been obtained from a distance-dependent
stress redistribution rule studied previously in
Ref.~\cite{Hidalgo2002}. One of the main quantities of interest in
this case is the critical stress above which the material breaks
down. In Ref.~\cite{Hidalgo2002}, a transition from a short- to a
long-range behavior has been observed numerically as a function of a
parameter characterizing the range of the redistribution. Within our
stochastic model, we were able to trace back this transition to a
change in the statistics of the stress redistribution.

The paper is organized as follows: In Sec.~\ref{sec:crdf},
we briefly discuss the calculation of current redistribution factors
and, in particular, refer to an interesting and very useful relation
that expresses $\Delta_{ij,mn}$ in terms of two-point resistances
$R_{kl}$. In the rest of this article, we will then study various
prototypical network topologies of growing complexity. In
Sec.~\ref{sec:lattice}, we will present our analytical and numerical
results for the behavior of the current redistribution in regular
one-, two-, and three-dimensional networks. As any failure in a
one-dimensional chain trivially leads to the breakdown of the entire
current flow, we consider in Sec.~\ref{sec:quasi_1d} the more
interesting case of quasi-one-dimensional networks (ladders). The
behavior of the current redistribution in small-world networks,
finally, is analyzed in detail in Sec.~\ref{sec:small_world}, and our
main conclusions are summarized in Sec.~\ref{sec:conclusions}.

\section{Computation of Current-Redistribution Factors}

\label{sec:crdf}

In the following, we consider a resistor network consisting of a set of
nodes~$i$ connected by bonds~$i$--$j$ with resistances~$r_{ij}>0$. At every
node, we allow for current injection or extraction, which is described by a
vector~$\mathbf{I}^\mathrm{s}$ with components $I^\mathrm{s}_i$, where
$I^\mathrm{s}_i>0$ $(<0)$ for injection (extraction) or zero otherwise. Obviously,
due to current conservation, the total current injection in every connected
part of the network has to vanish. Here, for simplicity, we consider the case of 
a connected network and, thus, require that
\begin{equation}
\label{eq:total_injection}
\sum_i I^\mathrm{s}_i=0\,,
\end{equation}
and also exclude the trivial case~$\mathbf{I}^{s}=\mathbf{0}$.

For the calculation of the current-redistribution
factors~$\Delta_{ij,mn}$ defined by Eq.~\eqref{eq:LRD}, we also need
to consider the perturbed network resulting from the removal of a
bond~$m$--$n$.  In order for Eq.~\eqref{eq:total_injection} to be
fulfilled for these perturbed networks, we have to assume that the
network remains singly connected after an arbitrary bond removal.

\label{sec:delta_via_kirchhoff}

According to Eq.~\eqref{eq:LRD}, we write
\begin{equation}
  \label{eq:delta_def}
 \Delta_{ij,mn} = (I_{ij}' - I_{ij})/I_{mn} \,,
\end{equation}
where the electrical currents $I_{ij}$, $I_{mn}$ and $I'_{ij}$ are determined by
solving the linear system of equations
\begin{equation}
  \label{eq:kirchhoff}
  \mathbf{I}^\mathrm{s} = \mathsf{Y} \, \mathbf{U}\,,
\end{equation}
following from Kirchhoff's circuit laws.
Here, $U_i$ is the voltage at node~$i$, and the nodal admittance matrix~$\mathsf{Y}$ is
given by
\begin{equation}
  \label{eq:Y_matrix}
  Y_{ij} = 
  -\frac{1}{r_{ij}} \quad (i \ne j) \quad\text{and}\quad Y_{ii} = -\sum_{k\ne i} Y_{ki}
\end{equation}
where $r_{ij}$ is the resistance of bond $i$--$j$.
From the solution of Eq.~\eqref{eq:kirchhoff}, one obtains the current
$I_{ij}$ across bond $i$--$j$ as
\begin{equation}
  \label{eq:current_from_voltage}
  I_{ij} = \frac{U_i - U_j}{r_{ij}}\,.
\end{equation}
Note that in accordance with Eq.~\eqref{eq:total_injection}, $\sum_i
Y_{ij} = 0$ for all nodes~$j$.  Thus, the matrix~$\mathbf{Y}$ does not
have full rank. In the case of a singly connected network considered
here, this means that the voltages $U_i$ are only determined up to an
arbitrary constant, which, however, drops out when calculating the
currents~\eqref{eq:current_from_voltage}. In order to determine the
current-redistribution factors by means of Eqs.~\eqref{eq:kirchhoff} and
\eqref{eq:current_from_voltage}, these equations have to be solved for
the unperturbed network and for each network resulting from the outage
of a bond~$m$--$n$.

Alternatively, the current-redistribution factors
can be obtained by the relation (for a derivation see
Appendix~\ref{sec:app_delta})
\begin{equation}
  \label{eq:Delta_from_R}
  \Delta_{ij,mn} = \frac{1}{2 r_{ij}} \frac{R_{in} - R_{im} + R_{jm} - R_{jn}}{1 - R_{mn} / r_{mn}}  
\end{equation}
where the two-point resistances between two arbitrary nodes~$k$ and $l$,
\begin{equation}
  \label{eq:two_point_R}
  R_{kl} = (U_k - U_l) / I_0\,,
\end{equation}
are determined by solving Eq.~\eqref{eq:kirchhoff}  for the unperturbed
network with 
\begin{equation}
  \label{eq:current_injections}
I^\mathrm{s}_k = - I^\mathrm{s}_l = I_0\,\text{, and } I^\mathrm{s}_{k'} = 0 \text{ for $k' \ne k$ or $l$}.  
\end{equation}
The $R_{kl}$ can  also be determined from the pseudo-inverse~$X$ of the nodal
admittance matrix~$Y$, see Eq.~\eqref{eq:R_from_X} of
Appendix~\ref{sec:app_delta}.  As there exist very efficient methods for the
calculation of~$X$ (see Ref.~\cite{Schaub2014}), the relation of
Eq.~\eqref{eq:Delta_from_R} offers a much more convenient way to determine the
$\Delta_{ij,mn}$ than the direct method described by Eqs.~\eqref{eq:delta_def}
to \eqref{eq:current_from_voltage} above.

The representation of Eq.~\eqref{eq:Delta_from_R} further shows
that the $\Delta_{ij,mn}$ are independent of the current
injections~$\mathbf{I}^\mathrm{s}$. They are thus only determined by the
topology of the network and by the values of the bond resistances. Note
that the sign of $\Delta_{ij,mn}$ depends on the \textit{a priori} arbitrary definition of the
current directions, i.e., 
\begin{equation}
\label{eq:delta_exch}
\Delta_{ji,mn} = -
\Delta_{ij,mn} \quad\text{and}\quad\Delta_{ij,nm} = - \Delta_{ij,mn}\,.
\end{equation}
We further note that $|\Delta_{ij,mn}| \le 1$.

\section{Regular $d$-Dimensional Lattices}

\label{sec:lattice}

\subsection{1d Chain}

\label{sec:1dchain}

For a 1d chain of resistors, the calculation of $\Delta_{ij,mn}$
depends on the chosen boundary conditions.
In the case of \textit{periodic boundary conditions}, i.e., for a \textit{1d ring} with $N$~identical resistors~$r$, we have
\begin{equation}
  \label{eq:R_1d}
  R_{ij} = R(d) = r d (1 - d/N) \,,\quad   d = |x_i - x_j|\,,
\end{equation}  
and thus
\begin{equation}
  R(d) = r d \,,\quad N\to\infty.
\end{equation}
From Eqs.~\eqref{eq:Delta_from_R} and \eqref{eq:R_1d}, it then follows that
\begin{equation}
        |\Delta_{ij,mn}| = |\Delta(d)| = 1 \text{ for all bonds $i$--$j$ and $m$--$n$}\,,
\end{equation}
where the sign of $\Delta_{ij,mn}$ depends on the relative ``direction''
of the two bonds along the ring [cf.\ Eq.~\eqref{eq:delta_exch}].

For a 1d chain with \textit{free boundary conditions}, on the other
hand, $\Delta_{ij,mn}$ is not properly defined by
Eq.~\eqref{eq:Delta_from_R}.

\subsection{2d Square Lattices}

\label{sec:2dlattice}

\subsubsection{Distance Dependence of $\Delta_{ij,mn}$}

For an infinite 2d square lattice of identical resistors~$r$, the two-point
resistances have been calculated, e.g., by Cserti~\cite{Cserti2000}.
In particular, Cserti determined the asymptotic form of 
the resistance~$R(x,y)$ between the origin $(0, 0)$ and node $(x,
y)$, in the limit of large $x$ and/or $y$,
\begin{equation}
  R(x,y) \simeq \frac{r}{\pi} 
  \left[
    \ln\sqrt{x^2 + y^2} + \gamma + \frac{\ln 8}{2}
  \right]\,,
  \label{eq:2d_R_asymp}
\end{equation}
where $\gamma=0.5772\dots$ is the Euler-Mascheroni constant.

\begin{figure}[ht]
  \centering
  \includegraphics[width=0.9\linewidth]{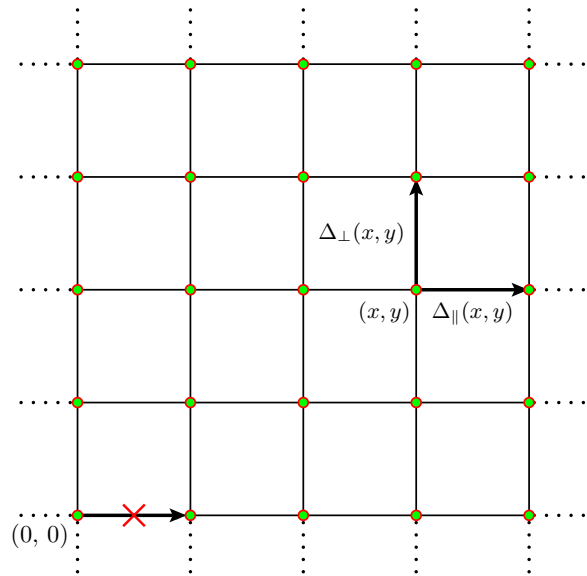}
  \caption{Definition of current-redistribution
  factors~$\Delta_\parallel(x,y)$ and $\Delta_\perp(x,y)$ in a 2d square
  lattice.}
  \label{fig:2d_square_lattice}
\end{figure}
To calculate the current-redistribution factors~$\Delta_{ij,mn}$, we
can, without loss of generality,
fix the failing bond $m$--$n$ to $m=(0,0)$ and $n=(1,0)$, see
Fig.~\ref{fig:2d_square_lattice}, and denote the $\Delta_{ij,mn}$-values
as $\Delta_\parallel(x,y)$ if $i$--$j$ is parallel to $m$--$n$, i.e.,
$i=(x,y)$ and $j=(x+1,y)$, or by $\Delta_\perp(x,y)$ for $i=(x,y)$ and
$j=x, y+1$. Introducing polar coordinates
\begin{equation}
  d = \sqrt{x^2 + y^2}\,,\quad\varphi = \arctan(y/x)\,,
\end{equation}
Eqs.~\eqref{eq:2d_R_asymp} and \eqref{eq:Delta_from_R} then lead to
asymptotic expressions
\begin{subequations}
  \begin{equation}
  \label{eq:2d_Delta_par}
  \Delta_\parallel(d,\varphi) \simeq 
    \frac{-1}{\pi d^2} \, \cos 2\varphi
\end{equation}
and
\begin{equation}
  \label{eq:2d_Delta_perp}
  \Delta_\perp(d,\varphi) \simeq -
    \frac{1}{\pi d^2} \, \sin 2\varphi + \dots\ .
\end{equation}
\label{eq:2d_Delta}
\end{subequations}
In addition to the asymptotic form of Eq.~\eqref{eq:2d_R_asymp},
Cserti~\cite{Cserti2000} also derived some recursion relations, from
which one can, in principle, calculate the $R(x,y)$-values exactly for
arbitrary nodes $x$ and $y$. A table of exact $R(x,y)$-values for
$0\le x,y\le20$ has been compiled by S.\ and R.\ Hollos~\cite{Hollos2005_Ref3}.
Using these values, Eq.~\eqref{eq:Delta_from_R} then allows us to
calculate exact values for $\Delta_\parallel(x,y)$ and
$\Delta_\perp(x,y)$. A few examples are given in Appendix~\ref{sec:app_2d_square},
Table~\ref{tab:2d_Delta}, and we also note that the recursion relations of Ref.~\cite{Cserti2000} lead
to the general result
\begin{equation}
  \label{eq:2d_Delta_diag}
  \Delta_\parallel(x,x) = 0\,.
\end{equation}

\begin{figure}[ht]
  \centering
  \includegraphics[width=0.95\linewidth]{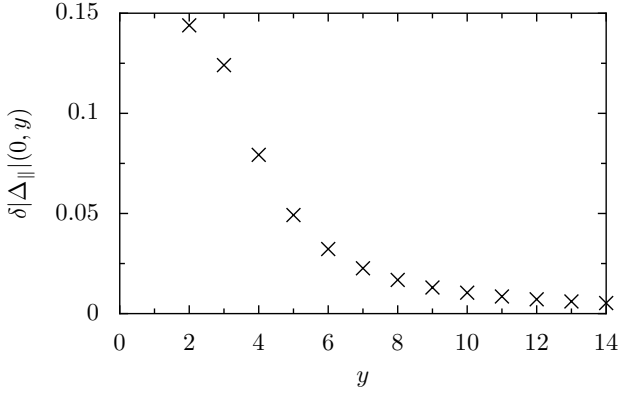}
  \caption{Infinite 2d square lattice: Comparison of exact values for $|\Delta_\parallel|(0,y)$ with
  their asymptotic approximations, $\delta|\Delta_\parallel| =
  \left[|\Delta_\parallel|^\mathrm{exact}
  -|\Delta_\parallel|^\mathrm{asymp}\right]/|\Delta_\parallel|^\mathrm{exact}$.}
  \label{fig:2d_comp_Delta}
\end{figure}

A comparison of exact and asymptotic values,
Eq.~\eqref{eq:2d_Delta_par}, for $|\Delta_\parallel|(0, y)$ is shown in
Fig.~\ref{fig:2d_comp_Delta}. It can be seen that the difference between
the exact and asymptotic values is smaller than $1\%$ if $y>10$.

\subsubsection{Statistics of $|\Delta|$-Values}

Motivated by our analysis of stochastic load-redistribution models for
cascading failure propagation~\cite{Lehmann2010}, we are also interested
in the overall statistics of the $|\Delta|$-values. To determine the
corresponding distribution function, we use a continuum approximation
similar to that developed in Ref.~\cite{Lehmann2010a}. We assume that
the failed element is located in the center of a circular ring  with an
inner radius of $d_\mathrm{min}$ and and outer radius of
$d_\mathrm{max}$. Assuming that the elements affected by the failure are
uniformly distributed in this circular ring, and using the 
asymptotic approximations of Eqs.~\eqref{eq:2d_Delta_par} and
\eqref{eq:2d_Delta_perp}, respectively, for $|\Delta_\parallel|$ and
$|\Delta_\perp|$, the probability distribution function for $|\Delta|$
can then be expressed in the form
\begin{equation}
\varrho(|\Delta|) = \frac{8}{\pi
(d_\mathrm{max}^2-d_\mathrm{min}^2)}
\int\limits_{d_\mathrm{min}}^{d_\mathrm{max}}\!\!\! \mathrm{d}\xi \xi \!\!
\int\limits_{0}^{\pi/2} \!\! \mathrm{d}\varphi \,\delta\left(\frac{\cos
2\varphi}{\pi \xi^2} -
|\Delta|\right)\,,
\label{eq:2d_Delta_cont}
\end{equation}
where we note that to account for both $|\Delta_\parallel|$ and
$|\Delta_\perp|$, an extra factor of $2$ has been included in the
prefactor. 

Carrying out the the integrations in Eq.~\eqref{eq:2d_Delta_cont}, we
obtain
\begin{multline}
\varrho(|\Delta|) = \frac{2}{\pi^2
(d_\mathrm{max}^2-d_\mathrm{min}^2) |\Delta|^2}
\bigg[
\sqrt{\smash[b]{1-\pi^2 d_\mathrm{min}^4 |\Delta|^2}}\\
-
\sqrt{1-\pi^2 \tilde d_\mathrm{max}^4 |\Delta|^2}
\bigg]\,,
\label{eq:2d_Delta_cont2}
\end{multline}
where in order to account for the $\delta$-function contraint in the
$\varphi$-integration, we have introduced the modified outer radius
$\tilde d_\mathrm{max} = \min(d_\mathrm{max}, 1/\sqrt{\pi\Delta})$.

As we assume that the continum approximation refers to a square lattice
of identical resistors with lattice constant equal to one, we set
$d_\mathrm{min}=1$ and determine $d_\mathrm{max}$ by the requirement
\begin{equation}
\pi (d_\mathrm{max}^2 - d_\mathrm{min}^2) = N\,,
\end{equation}
where~$N$ denotes the number of resistors (bonds) in the circular ring
between $d_\mathrm{min}$ and $d_\mathrm{max}$. We thus have
\begin{equation}
   d_\mathrm{max}  =
       \sqrt{\frac{N}{\pi} + 1}
   \label{eq:2d_rmax}
\end{equation}
and consequently 
\begin{equation}
   \tilde d_\mathrm{max}  =
        \min \left( \frac{1}{\sqrt{\pi |\Delta|}}, \sqrt{\frac{N}{\pi} +1}
        \right)\,.
   \label{eq:2d_tilde_rmax}
\end{equation}
Inserting Eqs.~\eqref{eq:2d_rmax} and \eqref{eq:2d_tilde_rmax} into
Eq.~\eqref{eq:2d_Delta_cont2}, we obtain (note that $d_\mathrm{min}=1$)
\begin{widetext}
\begin{equation}
\varrho(|\Delta|) = \frac{2}{\pi^2
N  |\Delta|^2}
\begin{cases}
\sqrt{1-\pi^2 |\Delta|^2}
-
\sqrt{1-\pi^2 (1+ N/\pi)^2 |\Delta|^2} &, \quad \displaystyle N|\Delta|\le\frac{1}{1+\pi/N}
\\
\sqrt{1-\pi^2 |\Delta|^2} &,\quad \displaystyle \frac{1}{1+\pi/N}
\le N |\Delta| \le \frac{N}{\pi}\,.
\end{cases}
\label{eq:2d_Delta_cont3}
\end{equation}
\end{widetext}

In order to be able to perform the limit~$N\to\infty$ in a meaningful
way, we introduce the scaled variable
\begin{equation}
  \label{eq:hat_Delta}
  \hat\Delta = N |\Delta|
\end{equation}
In the limit $N\to\infty$, the probability density for $\hat\Delta$ then becomes
a power law with a regularization below $\hat\Delta=1$
\begin{equation}
  \label{eq:2d_hat_Delta_rho}
  \varrho(\hat\Delta) = 
  \frac{2}{\pi \hat\Delta^2}
  \begin{cases}
    0 & \hat\Delta \le 0 \\
    1-\sqrt{1-\hat\Delta^2}    & 0<\hat\Delta\le1\\
    1 & \hat \Delta > 1\\
  \end{cases}
\end{equation}
and the corresponding cumulative distribution function has the form
\begin{equation}
    \mathrm{cdf}(\hat\Delta) = \frac{2}{\pi}
    \begin{cases}
      0 & \hat\Delta \le 0 \\
      \displaystyle
      \arcsin \hat\Delta - 
      \frac{1-\sqrt{1-\hat\Delta^2}}{\hat\Delta}
      &
      0 < \hat\Delta \le 1 \\
      \displaystyle
      \frac{\pi}{2}-\frac{1}{\hat\Delta}
      &
      \hat\Delta > 1
    \end{cases}
  \label{eq:2d_hat_Delta_cdf}
\end{equation}

From Eq.~\eqref{eq:2d_Delta_cont3}, we can further calculate the mean
value $\langle|\Delta|\rangle$, and we obtain
\begin{equation}
  \langle |\Delta| \rangle \simeq \frac{2}{\pi} \frac{\ln N}{N},\quad
  N\to\infty\,.
\end{equation}
for asymptotically large~$N$.

Finally, we can also consider the scaling of $|\Delta|$ with the
distance~$d$ from the failed resistor. The form of Eqs.~\eqref{eq:2d_Delta_par} and
\eqref{eq:2d_Delta_perp} motivates a consideration of the quantity
\begin{equation}
  \label{eq:tilde_Delta}
   \tilde\Delta = \pi d^2 |\Delta|\,
\end{equation}
and we find that this random variable is
distributed according to the density
\begin{equation}
  \varrho(\tilde\Delta) = 
  \begin{cases}
    \displaystyle
    \frac{2}{\pi\sqrt{1-\tilde\Delta^2}} & 0\le\tilde \Delta\le1\\
    0 & \text{otherwise.}
  \end{cases}
\end{equation}
The corresponding cumulative distribution function is then given by
\begin{equation}
  \mathrm{cdf}(\tilde\Delta) = \frac{2}{\pi}\arcsin \tilde\Delta\,,
  \label{eq:2d_tilde_Delta_cdf}
\end{equation}
where $\tilde\Delta$ varies in the range $0\le\tilde\Delta\le 1$.

\subsubsection{Numerical Results for Finite-Size Square Lattices}

We have compared the above analytical results, in particular Eqs.~\eqref{eq:2d_Delta},
\eqref{eq:2d_hat_Delta_cdf} and \eqref{eq:2d_tilde_Delta_cdf}, 
with numerical results for a finite-size square lattice of
dimensions $96\times96$ sites connected by $18'432$ bonds, where we have
assumed periodic boundary conditions in both $x$- and $y$-direction. 

\begin{figure*}[ht]
  \centering
  \includegraphics[width=0.95\textwidth]{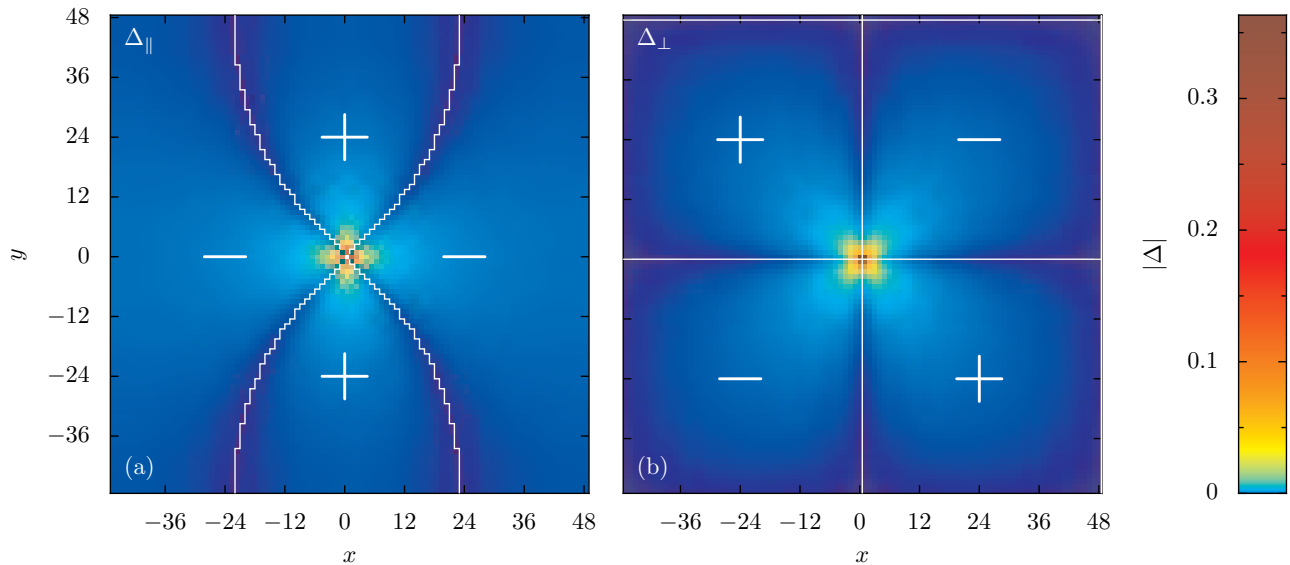}
  \caption{Current-redistribution factors in a 2d square lattice consisting of
    $96\times96$ sites and assuming periodic boundary conditions. The failing
    bond is oriented horizontally in the center of the graph, i.e. connects
    sites $(0,0)$ and $(1,0)$. The left (right) panel shows the
    current-redistribution factors in bonds parallel, $\Delta_\parallel$,
    (perpendicular, $\Delta_\perp$) to the failing bond. Regions of positive
    and negative~$\Delta$ are indicated.}
  \label{fig:2d_square_Delta}
\end{figure*}
Figure~\ref{fig:2d_square_Delta} shows the dependence of the
current-redistribution factors on orientation and position with respect to the
failing bond. In the vicinity of the failing bond, we observe that (as for the
infinitely large grid [cf.\ Eq.~\eqref{eq:2d_Delta_diag}]) the
current-redistribution factors~$\Delta_\parallel$ for bonds parallel to the
failing bond vanish along the diagonals. For larger distances from the
failing bond, we observe finite-size effects which lead to a deviation from
this behavior. It can be shown that the deviations are due to the fact that
the current flows at the boundaries of the finite-size square lattice differ
considerably from those in an infinite square lattice. We also note that the
form of the deviations observed in Fig.~\ref{fig:2d_square_Delta}(a) depends
strongly on the chosen boundary conditions (e.g., periodic vs.\ free boundary
conditions).

\begin{figure}[ht]
  \centering
  \includegraphics[width=0.95\linewidth]{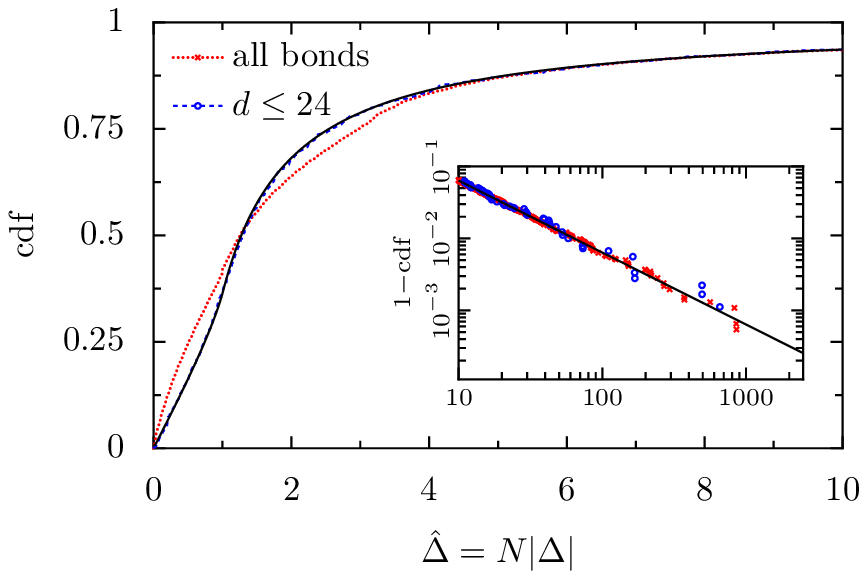}
  \caption{Cumulative distribution function of the size-scaled magnitude of the
    current-redistribution factors~$\hat \Delta = N |\Delta|$ for a 2d square
    lattice of $96\times96$ sites assuming periodic boundary
    conditions. Dotted (red) line: result for all bonds. Dashed (blue) line:
    only bonds within a distance $d\le24$ of the
    failing bond. The analytical result~\eqref{eq:2d_hat_Delta_cdf} for an
    infinitely large system within a continuum approximation is indicated by
    the solid line. The inset shows the tail of the
    distribution function for large~$\hat\Delta$. Crosses (red): all
    bonds. Squares (blue): only bonds within a distance $d\le24$ of the
    failing bond }
  \label{fig:2d_square_DeltaN}
\end{figure}
Finite-size effects can also be observed in the statistics of the
current-redistribution factors. In Fig.~\ref{fig:2d_square_DeltaN}, we plot
the empirical cumulative distribution function of the size-scaled magnitude of the
current-redistribution factors~$\hat\Delta = N |\Delta|$
[Eq.~\eqref{eq:hat_Delta}]. When comparing with the analytical
result~\eqref{eq:2d_hat_Delta_cdf} for the case of an infinitely large system
within a continuum approximation (solid line), we find a very good agreement
if we restrict the statistics to the bonds within a distance~$d\le24$ (blue
circles) to the failed bond. Otherwise, deviations for very small
$|\Delta|=\mathcal{O}(1/N)$ can be observed. For the upper tail of the
distribution function shown in the inset, i.e., for the large $|\Delta|$
values of the order~$1$, such deviations cannot be found.

\begin{figure}[ht]
  \centering
  \includegraphics[width=0.95\linewidth]{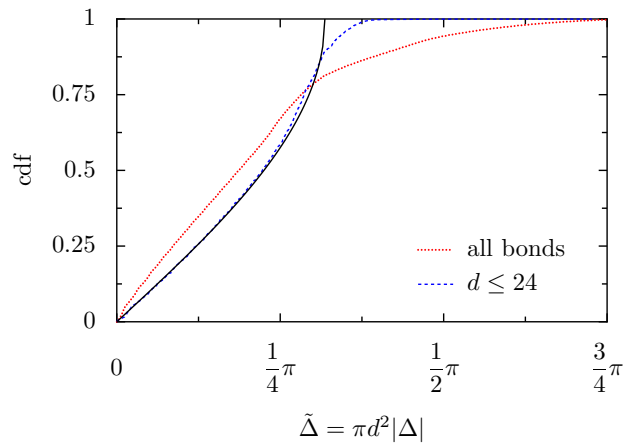}
  \caption{Cumulative distribution function of the distance-scaled magnitude
    of the current-redistribution factors~$\tilde \Delta = \pi d^2 |\Delta|$
    for a 2d square lattice of $96\times96$ sites assuming periodic boundary
    conditions. The meaning of the symbols is as in
    Fig.~\ref{fig:2d_square_DeltaN} and the solid line denotes the analytical
    result~\eqref{eq:2d_tilde_Delta_cdf}.}
  \label{fig:2d_square_Deltar}
\end{figure}
Finally, the statistics of the distance-scaled magnitude of the
current-redistribution factors~$\tilde \Delta = \pi d^2 |\Delta|$
[Eq.~\eqref{eq:tilde_Delta}] is shown in Fig.~\ref{fig:2d_square_Deltar}.
Here, the distance-scaling leads to a more pronounced appearance of the
finite-size effects, but again the statistics of the current-redistribution
factors for the bonds in the vicinity of the failed bond ($d\le24$) is well described by
the simple analytical result of Eq.~\eqref{eq:2d_tilde_Delta_cdf} from the
continuum approximation (solid line).

\subsection{3d Simple Cubic Lattice}

For a infinite 3d simple cubic lattice of identical unit resistors, the two-point resistances $R(x, y, z)$ can be
expressed in terms of the Green functions $G(x, y, z; 3)$, 
see e.g. Ref.~\cite{Cserti2000},
\begin{equation}
  \begin{split} 
    R(x, y, z) & = G(0, 0, 0; 3) - G(x, y, z; 3)\\ & = R(\infty) - G(x, y, z; 3)\,.
    \end{split}
\end{equation}
In Ref.~\cite{Joyce2002}, it is shown that
\begin{equation}
    \begin{split} 
      R(\infty) & = \frac{1}{96\pi^3} (\sqrt{3}-1) \big[\Gamma(1/24)\Gamma(11/24)\big]^2
      \\
      &=0.505462\dots
    \end{split}
\end{equation}
and that $G(x, y, z; 3)$ has the following asymptotic expansion,
\begin{equation}
  \label{eq:Gasymp3d}
  \begin{split}
  G(x,y,z; 3) \simeq \frac{1}{2\pi d} & + 
   \frac{1}{8\pi d^7}\big[ (x^4+y^4+x^4) \\&- 3(x^2 y^2 + x^2 z^2 + y^2 z^2)\big] 
  \end{split}
\end{equation}
for $d=\sqrt{x^2+y^2+z^2}\to\infty$.

We now consider the removal of bond $m$--$n$, where $m = (0, 0, 0)$ and
$n = (0, 0, 1)$. The current-redistribution factors
$\Delta_\parallel^{(z)} (x,y,z)$ for affected bonds parallel to the
$z$-axis can then be expressed as [see Eq.~\eqref{eq:Delta_from_R}]
\begin{equation}
        \Delta_\parallel^{(z)}(x,y,z) = \frac{ R (x,y,z+1) + R(x,y,z-1) -2 R(x,y,z)}{2[1-R(0,0,1)]}\,.
\end{equation}

Using Eq.~\eqref{eq:Gasymp3d} and $R(0, 0, 1) = 1/3$, we then obtain
\begin{subequations}
  \begin{equation}
  \begin{split}
  \Delta_\parallel^{(z)}(x,y,z) & {} 
  \simeq \frac{3}{8\pi d^3 } [1-3(z/d)^2 ]
  \\
  & =\frac{3}{8\pi d^3} [1-3 \cos^2\theta]
  \end{split}
\end{equation}
for $d=\sqrt{x^2+y^2+z^2}\to\infty$, where $\theta$ denotes the angle between the $z$-axis and
the vector $(x, y, z)$.

Similarly, for affected bonds parallel to the
$x$-($y$-)axis, we obtain
\begin{equation}
  \Delta_\perp^{(z)}(x,y,z) 
  \simeq  
 - \frac{9}{16\pi d^3} \sin 2\theta \cos \varphi\,,
\end{equation}
where $\varphi$ refers to the angle between the $x$-($y$-)axis and the
vector $(x,y,0)$, respectively.
\end{subequations}

Apart from the angular dependences, the dominating asymptotic behavior of
$|\Delta|$ is thus given by
\begin{equation}
  |\Delta| \simeq \frac{1}{d^3}\,
\end{equation}
compared with the $1/d^2$ behavior of $|\Delta|$ for a 2d square lattice
[cf.\ Eqs.~\eqref{eq:2d_Delta_par} and \eqref{eq:2d_Delta_perp}].

Using the same continuum approximation as in Sec.~\ref{sec:2dlattice}, we can 
show that the dominating behavior of the distribution
function~$\varrho(|\Delta|)$ is given by the power law
\begin{equation}
  \varrho(|\Delta|) \simeq \frac{1}{|\Delta|^2}\,.
\end{equation}
Note that this behavior is identical to that of 2d regular lattices,
see Eq.~\eqref{eq:2d_hat_Delta_rho}, and we expect that this 
is also valid for all regular lattices of dimension four and higher.  For
quasi-one-dimensional lattices, as well as for small world networks,
however, the behavior of $|\Delta|(d)$ and of $\varrho(|\Delta|)$ is
completely different, as we shall see in the following sections.

\section{Quasi-One-Dimensional Networks}

\label{sec:quasi_1d}

For a strictly one-dimensional system (1d chain) with free boundary
conditions, every bond removal leads to a separation of the network
into two unconnected parts, so that the current-redistribution factors
are not defined.

\begin{figure}[ht]
  \centering
  \includegraphics[width=0.95\linewidth]{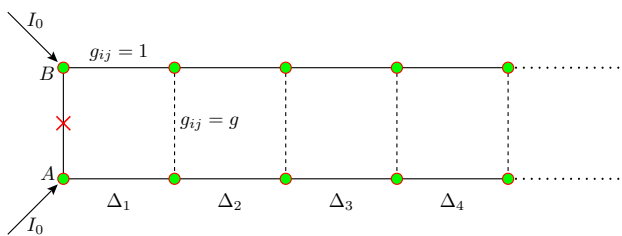}
  \caption{Semi-infinite ladder consisting of bonds with unit conductances along the ladder as
    well as for the leftmost spoke (solid line), which is assumed to fail, and equal
    conductances $g$ of the other spokes (dashed line).}
  \label{fig:ladder}
\end{figure}

We can, however, consider quasi-one-dimensional systems such as, e.g.,
the semi-infinite ladder shown in Fig.~\ref{fig:ladder}. Here, we have
assumed unit conductances, $g=1$, along the ladder as well as for the
leftmost spoke and equal conductances~$g$ for all other spokes.

We then restrict ourselves to a failure of the leftmost spoke and consider the
current-redistribution factors~$\Delta_n$ in an infinite (one-sided) ladder
for the bonds along the ladder (see Fig.~\ref{fig:ladder}). An 
analytical calculation of the current-redistribution factors~$\Delta_n$ is
straightforward and leads to the result
\begin{equation}
  \Delta_n = X(g)^{-(n-1)}\,
\end{equation}
where
\begin{equation}
  \label{eq:ladder_Xg}
  X(g) = \frac{G+g}{G} = 
  \frac{\sqrt{g^2+2g}+g}{\sqrt{g^2+2g}-g}\,.
\end{equation}
The current-redistribution factors~$\Delta_n$ thus depend exponentially on the
distance $d=n$ from the removed bond. The slope~$s(g)$ of the linear
relation~$\ln \Delta_n$ vs.~$n$ is given by
\begin{equation}
  \label{eq:ladder_sg}
  s(g) = \ln X(g)\,.
\end{equation}
For $g\to0$, this slope becomes $s(g) \approx \sqrt{2g}$, and in the case
where all conductances are equal, i.e., $g=1$, we have 
\begin{equation}
  s(1) = \ln\left(\frac{\sqrt{3}+1}{\sqrt{3}-1}\right)\,.
\end{equation}

\section{Ring topologies with Long-Range Connections}

\label{sec:small_world}

A particulary interesting class of networks can be obtained when random
long-range links are added to originally regular structures such as the
ones studied in the previous sections. In a seminal
paper~\cite{Watts1998}, Watts and Strogatz found that when only a small
fraction of bonds in a ring lattice are randomly rewired, the average
shortest path length drops significantly and reaches values like in a
completely random graph. The local structure, however, is retained.
Watts and Strogatz argued that such a semi-random model captures crucial
properties of many real world graphs including, e.g., power grids.
Additionally, by the nature of their construction, such networks (and
their generalization to higher dimensions) maintain the underlying
spatial structure including its (Euclidean)
metric~\cite{Barthelemy2011}. In the context of our investigations, this
will allow us to study the distance dependence of the current
redistribution also for non-regular networks that represent more
realistic models for infrastructure networks. We further note that our
statistical approach is a particularly adequate method to study current
redistribution in this type of networks, as they are random by
construction. Care has to be taken, however, that now in addition to the
statistics of the affected bonds, the disorder statistics of the network
itself has to be considered.

In this Section, we will first consider the behavior of the current
redistribution in the case of a randomly placed single long-range
connection in a ring. As a second extreme case we will then look at a
ring with additional connections added between each non-neighbouring
pair of nodes, i.e. a completely connected graph. Finally, we will
present extensive numerical calculations for the current redistribution
in a variant of the Watts-Strogatz model where, instead of rewiring
connections, additional long-range shortcuts are randomly
added~\cite{Newman1999a}.

%\subsection{One-Dimensional Ring without Long-Range Connections}

%As already discussed in section~\ref{sec:1dchain}, we have the obvious
%result that $|\Delta_{ij,mn}|= |\Delta(d)| = 1$ for all bonds $i$--$j$ and $m$--$n$,
%i.e., the current-redistribution factors are all $1$ or $-1$.
%
\subsection{Ring with a Single Long-Range Connection}

\begin{figure}[ht]
  \centering
  \includegraphics[width=0.85\linewidth]{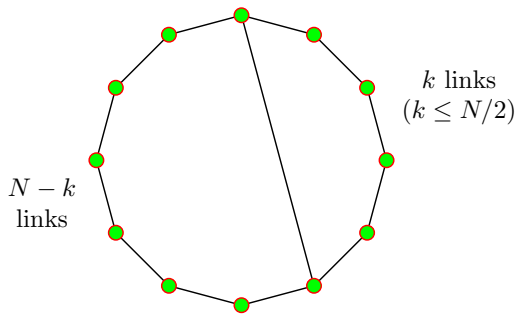}
  \caption{Ring with one long-range connection.}
  \label{fig:ring_lrc}
\end{figure}

The case of a ring with a single long-range connection can be treated
analytically.  Consider the graph shown in Fig.~\ref{fig:ring_lrc} and
assume that a bond in the segment with $k$~links fails. Then the
current-redistribution factors only assume two different values:
$|\Delta|=1$ if the considered bond is in the same segment as the
failed bond, and $|\Delta| = 1/(N-k+1)$ if it is in the other
segment. If a bond in the segment with $N-k$ links fails, the two
values are $|\Delta|=1$ and $|\Delta|=1/(k+1)$.

Using a straightforward counting procedure, we can easily determine
the frequencies with which the three $|\Delta|$-values occur when we
vary the position of the failed bond. From these we can then calculate
the average of $|\Delta|(d)$ with respect to all possible positions of
the failed bond, $\langle|\Delta|\rangle(d,k;N)$, where~$d$ refers to
the distance (measured along the ring) between the considered and the
failed bond, $k$ is explained in Fig.~\ref{fig:ring_lrc}, and $N$ is
the size of the ring (number of sites).

In the limit as $N\to\infty$, we can introduce the normalized variables $\delta=d/N$ and $\kappa=k/N$,
where $\delta$ and $\kappa$ can be considered as continuous ($0\le \delta, \kappa \le 1/2$). It follows that
 in this limit, $\langle|\Delta|\rangle$ can be written as
\begin{equation}
  \langle |\Delta| \rangle (\delta, \kappa) =
  \begin{cases}\displaystyle
    1- 2 \delta \,, & 0\le \delta\le \kappa\\
    \displaystyle
    1- 2 \kappa \,, & \kappa\le \delta\le 1/2\,,
  \end{cases}
\end{equation}
and its average over all normalized lengths~$\kappa$ of the long-range links becomes
\begin{equation}
  \label{eq:ring_lrc_doubavg1}
  \langle\!\langle |\Delta| \rangle\!\rangle (\delta) = 
  1- 2\delta + 2\delta^2\,,\quad 0\le\delta\le 1/2\,.
\end{equation}

Another quantity of interest that can be calculated analytically in the limit
as $N\to\infty$, is the graph-distance $\tilde\delta$ between two points on
the ring. The calculations are again straightforward but somewhat more tedious
than for $|\Delta|$, as we have to consider considerably more sub-cases. The
final result is
\begin{subequations}
  \begin{equation}
  \langle \tilde\delta \rangle (\delta, \kappa)
  =
  \begin{cases}
    \delta - \kappa \delta + \kappa^2/2\, & 0\le\kappa\le 2\delta \\
    \delta\,                              & 2\delta \le \kappa \le 1/2
  \end{cases}
\end{equation}
if $0\le\delta\le1/4$ and 
\begin{equation}
  \langle \tilde\delta \rangle (\delta, \kappa)
  =
  \begin{cases}
    \delta - \kappa \delta + \kappa^2/2\, & 0\le\kappa\le 1-2\delta \\
    1/2-\kappa + \kappa^2\,               & 1-2\delta \le \kappa \le 1/2
  \end{cases}
\end{equation}
\end{subequations}
if $1/4\le\delta\le1/2$. The average over all lengths~$\kappa$ of the the
long-range link finally becomes
\begin{equation}
  \label{eq:ring_lrc_tilde_delta}
  \langle\!\langle \tilde \delta \rangle\!\rangle =
  \delta - \frac{4}{3}\, \delta^3\,.
\end{equation}

\begin{figure}[ht]
  \centering
  \includegraphics[width=0.95\linewidth]{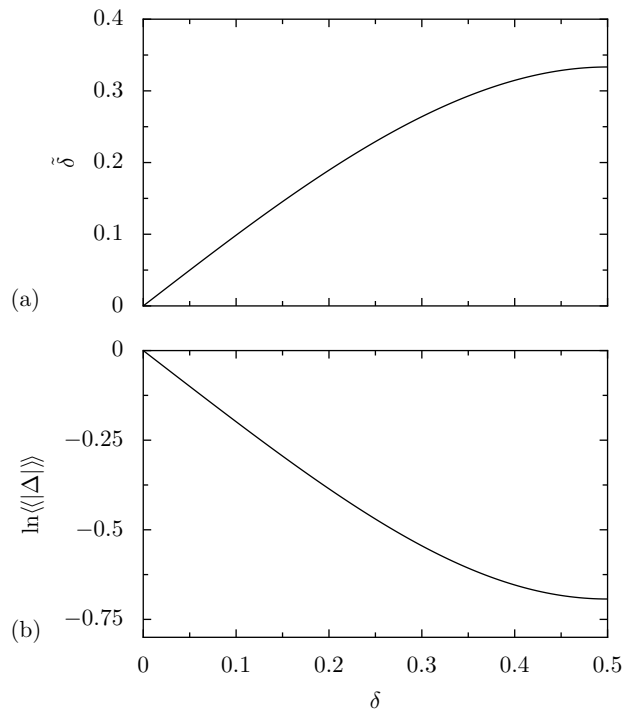}
  \caption{Behavior of (a) $\langle\!\langle \tilde\delta
  \rangle\!\rangle$ [Eq.~\eqref{eq:ring_lrc_tilde_delta}] and (b) $\ln\langle\!\langle |\Delta|\rangle\!\rangle$ [Eq.~\eqref{eq:ring_lrc_doubavg1}] vs.~$\delta$.}
  \label{fig:ring_lrc_dist_lnDelta}
\end{figure}
It is interesting to note the similarity in the behavior of $\langle\!\langle
\tilde\delta \rangle\!\rangle$ and $\ln\langle\!\langle |\Delta|
\rangle\!\rangle$ (see Fig.~\ref{fig:ring_lrc_dist_lnDelta}). Both quantities
start out linearly for $0\le \delta \approx 0.2$ and then saturate as
$\delta\to0.5$. This suggests that we have the following approximate relation
between $\langle\!\langle |\Delta|
\rangle\!\rangle$ and $\langle\!\langle
\tilde\delta \rangle\!\rangle$:
\begin{equation}
  \label{eq:ring_lrc_dist_Delta}
 \langle\!\langle |\Delta|\rangle\!\rangle 
 \approx 
 \mathrm{e}^{-\lambda \langle\!\langle \tilde\delta \rangle\!\rangle}\,.
\end{equation}
As in quasi-one-dimensional networks (Sec.~\ref{sec:quasi_1d}), we thus
observe at least an approximate exponential dependence of the
current-redistribution factors~$|\Delta|$ on the distance between the
considered and the failed bond, but here only in an average sense and on if we
replace the Euclidean distance along the ring by the graph distance.

The relation of Eq.~\eqref{eq:ring_lrc_dist_Delta} is surprisingly
accurate if we set $\lambda \approx 2$. The value $\lambda = 2$ is
obtained if we require that the exact and approximate expressions for
$\langle\!\langle|\Delta|\rangle\!\rangle$,
Eqs.~\eqref{eq:ring_lrc_doubavg1} and \eqref{eq:ring_lrc_dist_Delta},
respectively, have equal initial slopes (at $\delta=0$). If, on the other
hand, we require that Eq.~\eqref{eq:ring_lrc_dist_Delta} reproduces the
exact value of $\langle\!\langle|\Delta|\rangle\!\rangle = 1/2$ at
$\delta=1/2$, then $\lambda = 3\ln2=2.079$, which appears to give an
even superior approximation.

\subsection{Fully Connected Ring}

\label{sec:toconn}

We now consider the opposite extreme of complete connectivity, i.e., each site
on the ring is connected to every other site.

\begin{figure}[ht]
  \centering
  \includegraphics[width=0.95\linewidth]{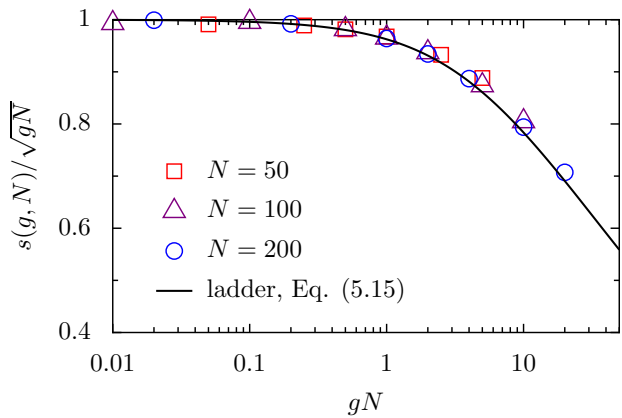}
  \caption{Behavior of $s(g,N)$, the exponent that characterizes the
  decrease of $|\Delta(d)|$ in a fully connected ring of size $N$ with
  long-range conductances~$g$. Comparison is made with the corresponding
  exponent of a one-dimensional ladder with spoke-conductances $gN/2$,
  see Eq.~\eqref{eq:toconn_sg}.}
  \label{fig:toconn_vs_ladder}
\end{figure}

If all connections have the same conductance, $g_{ij}=1$, it can be shown
that
\begin{equation}
  \begin{split}
  R_{ij} &= \frac{2}{N}\quad \text{for all $i\ne j$}\,,\\
  R_{ii} &= 0
  \end{split}
\end{equation}
This leads to 
\begin{equation}
  |\Delta_{ij,mn}| = 
  \begin{cases}
    \displaystyle \frac{1}{N-2}
    & \text{if bond $i$--$j$ is adjacent to $n$--$m$}\\
      0 & \text{otherwise.}
  \end{cases}
\end{equation}
The average value of $|\Delta|$ thus becomes
\begin{equation}
  \langle | \Delta | \rangle = \frac{2}{(N-1)(N-2)}
  \simeq \frac{2}{N^2}\quad(N\to\infty)\,.
\end{equation}
If the long-range links have smaller conductances than the nearest neighbor
links on the ring, i.e., $g^\mathrm{long range}_{ij} = g < 1$, we are no
longer able to determine the $|\Delta_{ij,mn}|$-values analytically.
Numerical simulations, however, show unambigously that $|\Delta(d)|$
decreases exponentially with the
distance~$d$ (measured along the ring) from the failed link,
\begin{equation}
  |\Delta(d)| \approx \mathrm{e}^{-s(g, N) d}
\end{equation}
with an exponent $s(g, N)$ that approaches the value
\begin{equation}
  s(g, N) \approx \sqrt{N g} \quad \text{if $Ng\ll 1$}\,.
\end{equation}
For the fully connected ring with long-range conductances~$g<1$,
$|\Delta(d)|$ thus exhibits quasi-one-dimensional behavior, and this
observation can even be further quantified: It turns out that in the
limit of large~$N$, a fully connected ring is very accurately
approximated by a one-dimensional ladder of size $N/2$ with spoke
conductances~$g^\text{spoke}=gN/2$, if $g\ll 1$, i.e.\ [see
  Eqs.~\eqref{eq:ladder_Xg} and \eqref{eq:ladder_sg}],
\begin{equation}
  \label{eq:toconn_sg}
  s(g, N) \approx 
  \ln 
  \frac{\sqrt{g^2+4g/N} + g}{\sqrt{g^2+4g/N}-g}\,,
  \quad\text{if $g\ll 1$ and $N\gg 1$}\,.
\end{equation}
The accuracy of the approximation~\eqref{eq:toconn_sg} is demonstrated by
the numerical results shown in Fig.~\ref{fig:toconn_vs_ladder}.

\subsection{Random Small-World Networks}

\label{sec:random_small_world}

Finally, we turn our attention to the case of random small-world
networks, which we construct from a (one-dimensional) ring by adding
random shortcuts similar to the model by Newman and Watts in
Ref.~\cite{Newman1999a}. Specifically we start from a ring of $N$
sites with each site initially being connected to its two nearest
neighbors and then add long-range connections randomly with
probability $2p/N$ between the $N(N-3)/2$ different pairs of
non-nearest neighbours, so that the average number of long-range links
is $p(N-3) \simeq p N$ for large~$N$~\footnote{In contrast to
  Ref.~\cite{Newman1999a} this construction avoids self-loops and
  multiple connections between the same nodes.}.

\begin{figure}[ht]
  \centering
  \includegraphics[width=\linewidth]{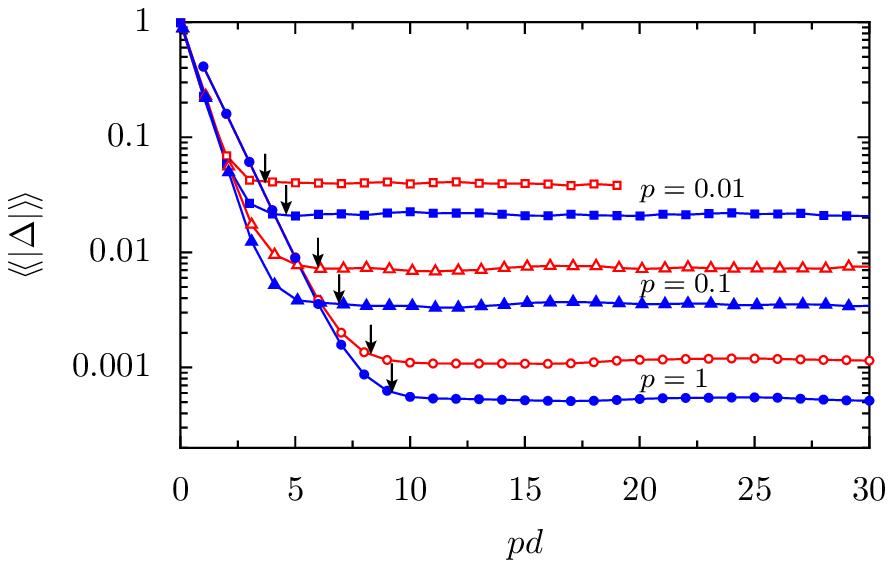}
  \caption{Averaged $|\Delta|$ as a function of the scaled ring
  distance~$p d$ for rings of size $N=4000$ (red, open symbols) and $10000$ (blue, closed symbols),
  and for different values of the probability parameter~$p$. The
  numerical results represent averages over both the location of the
  failing bond and $25$ realizations of the
  small-world network. The lines serve as a guide to the
  eye. The arrows indicate the location of the critical ring distances 
  $d_0(p,N)$ [Eq.~\eqref{eq:ring_dgraph}]}
  \label{fig:sw_Delta_mean_rd_scaled}
\end{figure}

\begin{figure}[ht]
  \centering
  \includegraphics[width=\linewidth]{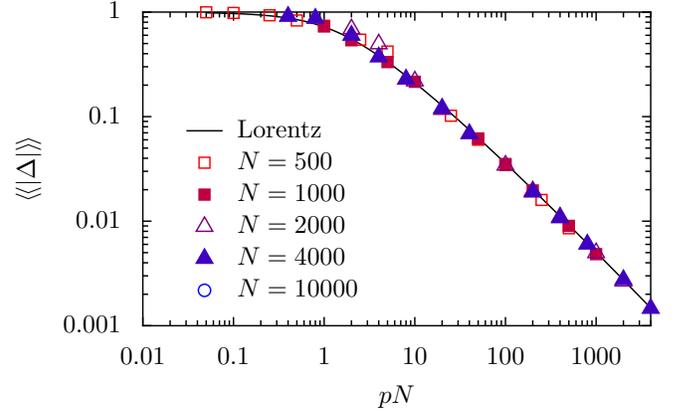}
  \caption{Average $\langle\!\langle|\Delta|\rangle\!\rangle$ of the
  current-redistribution factors $|\Delta_{ij,mn}|$ for rings of
  different size~$N$ and for different probabilities~$p$ of adding
  long-range connections. The
    numerical results represent averages over both the location of the
      failing bond and $25$ realizations of the
        small-world network. The line marked ``Lorentz'' corresponds to the
  approximation of Eq.~\eqref{eq:Delta_av_lorentz} for
  $\langle|\Delta|\rangle$ vs.~$pN$.}
  \label{fig:sw_Delta_mean}
\end{figure}

\begin{figure}[ht]
  \centering
  \includegraphics[width=\linewidth]{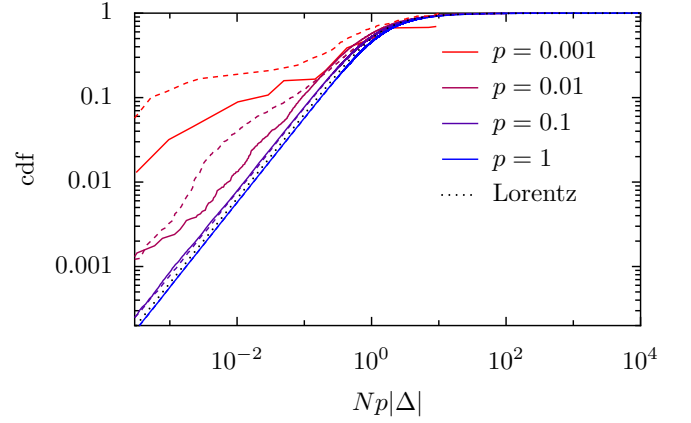}
  \caption{Cumulative distribution function of the scaled
    current-redistribution factors, $N p \langle|\Delta|\rangle$, for
    a ring with $N=10000$ sizes and different $p$-values for two
    different realizations of the random long-range links (solid and
    dashed lines). The dotted curve denoted by ``Lorentz'' corresponds
    to Eq.~\eqref{eq:cdf_lorentz}}
  \label{fig:sw_Delta_cdf}
\end{figure}

To investigate the behavior of $\Delta_{ij,mn}$ in these small-world
networks, we have performed a large number of simulations on networks
of varying sizes (up to $N=10000$) and for different values of~$p$
($0.001 \le p \le 1$). A selection of corresponding results is shown in
Figs.~\ref{fig:sw_Delta_mean_rd_scaled}--\ref{fig:sw_Delta_cdf}.

Figure~\ref{fig:sw_Delta_mean_rd_scaled} refers to the behavior of
$\langle\!\langle |\Delta| \rangle\!\rangle$ as a function of the scaled ring distance $p
d$. The average $\langle\!\langle|\Delta|\rangle\!\rangle$ is an average
over 25 different realizations of the corresponding random network as
well as an average over all locations of the failed bond and the
corresponding (two) affected connections at the given distance. We
note that failures and their effects are only considered for bonds on the ring.
We observe that the behavior of
$\langle\!\langle|\Delta|\rangle\!\rangle$ vs.~$p
d$ is distinctly different from that of regular networks:
For small values of $p d$, $\langle\!\langle|\Delta|\rangle\!\rangle$
initially decreases exponentially with $d$,
\begin{equation}
  \langle\!\langle|\Delta|\rangle\!\rangle \approx \exp(-\lambda p d)\,,
\end{equation}
i.e., the network behaves like a quasi-one-dimensional system. Around
some critical value of $d$, $d =
d_\mathrm{c}(p, N)$, however, $\langle\!\langle|\Delta|\rangle\!\rangle$
exhibits a
transition and stays constant for larger values of $p d$,
\begin{equation}
   \langle\!\langle|\Delta|\rangle\!\rangle \approx \Delta_0(p, N)
   \quad\text{if $d \gtrsim d_\mathrm{c}(p, N)$.}
\end{equation}

Interestingly, the behavior of $-\ln \langle\!\langle|\Delta|\rangle\!\rangle$
vs.~$d$ is remarkably similar to that of the graph distance
(shortest-path length) in a small-world network~\cite{Menezes2000},
\begin{equation}
  \label{eq:ring_dgraph}
  d_\mathrm{graph}
  \approx
  \begin{cases}
    \displaystyle
    d & \text{if $d \lesssim  d_0 = \ln(p N)/p$} \\
    d_0 & \text{if $d \gtrsim  d_0$}
  \end{cases}\,,
\end{equation}
and we note that the critical distance $p d_\mathrm{c}(p, N)$ above
which $\langle\!\langle|\Delta|\rangle\!\rangle$ saturates coincides quite well with the
critical value $p d_0 = \ln (pN)$ in Eq.~\eqref{eq:ring_dgraph}, see
Fig.~\ref{fig:sw_Delta_mean_rd_scaled}. From the above observations and
considerations, we may be tempted to express
$\langle\!\langle|\Delta|\rangle\!\rangle$ as
\begin{equation}
  \langle\!\langle|\Delta|\rangle\!\rangle
  \approx
  \exp(-\lambda p d_\mathrm{graph})\,.
  \label{eq:sw_Delta_0_approx_gen}
\end{equation}
It then turns out that this relation yields a very accurate description
of the plateaus $\Delta_0(p, N)$ if we set $\lambda=0.825$,
\begin{equation}
  \Delta_0(p, N) \approx
  \exp(-0.825 p d_0) = (p N) ^ {-0.825},
  \label{eq:sw_Delta_0_approx}
\end{equation}
see Table~\ref{table:sw_plateau}.
\begin{table}[t]
   \begin{tabular}{lll|lll}
       & & & \multicolumn{3}{c}{$\log_{10}(\Delta_0)$} \\
       $p$ & $N$ & $p N$ & Simulation & Eq.~\eqref{eq:sw_Delta_0_approx}
       & Eq.\eqref{eq:Delta_av_lorentz_large}
       \\
       \hline
       $0.01$ & $4000$  & $40$    & $-1.40$ & $-1.32$ & $-1.23$ \\
       $0.01$ & $10000$ & $100$   & $-1.67$ & $-1.65$ & $-1.53$ \\
       $0.1$  & $4000$  & $400$   & $-2.13$ & $-2.15$ & $-2.08$ \\
       $0.1$  & $10000$ & $1000$  & $-2.45$ & $-2.48$ & $-2.36$ \\
       $1$    & $4000$  & $4000$  & $-2.95$ & $-2.97$ & $-2.88$ \\
       $1$    & $10000$ & $10000$ & $-3.28$ & $-3.30$ & $-3.23$
   \end{tabular}
   \caption{Comparison of the observed plateau values~$\Delta_0(p, N)$
   with two analytical expressions, Eq.~\eqref{eq:sw_Delta_0_approx} and
   Eq.~\eqref{eq:Delta_av_lorentz_large}, see text.}
   \label{table:sw_plateau}
\end{table}
In this context, it may be worth noting that Korniss et al.~\cite{Korniss2006} have
derived an asymptotic expression for the two-point resistance in
small-world networks in the limit $N\to\infty$,
\begin{equation}
  \langle\!\langle R(d) \rangle\!\rangle
  \simeq
  1-\exp(-\sigma d)\,,
\end{equation}
and an analysis of their numerical results shows that these can be very
accucurately described by $\sigma = 0.825 p$.

For small values of $d_\mathrm{graph}$, however, where
$d_\mathrm{graph} = d$, expression~\eqref{eq:sw_Delta_0_approx_gen}
only approximates the observed behavior of $\langle\!\langle|\Delta|\rangle\!\rangle$ if
we assume that $\lambda$ becomes $p$-dependent. From our numerical
simulation (see, e.g.,~Fig.~\ref{fig:sw_Delta_mean_rd_scaled}), we
observe that $\ln \langle\!\langle|\Delta|\rangle\!\rangle$ vs.~$pd$ exhibits an
initial slope $\lambda(p) \approx 1.4$ if $p\lessapprox 0.1$,
$\lambda(p{=}0.2)\approx 1.28$, $\lambda(p{=}0.5)\approx 1.11$, and
$\lambda(p{=}1)\approx 0.93$. We also note that $\lambda(p)$ appears
to be independent of system size~$N$, but the observed $p$-dependence
of $\lambda(p)$ does not seem to have an obvious explanation.

In addition to the dependence of $\langle\!\langle|\Delta|\rangle\!\rangle$ on the
distance~$d$ between affected and failed bond, we have also analyzed
the overall statistics of the $|\Delta|$-values. Corresponding
numerical results are shown in Figs.~\ref{fig:sw_Delta_mean} and
\ref{fig:sw_Delta_cdf}.

A detailed analysis of these numerical results has shown that many of
the observed features can be described if we assume that the statistics
of $pN|\Delta|$-values is a half-Lorentz distribution (of width
$\gamma=1$) with a cutoff at $|\Delta|=1$ and a $\delta$-function at
$|\Delta|=1$, i.e.
\begin{equation}
  \label{eq:rho_lorentz}
  \varrho(y) = \frac{2}{\pi} \, \frac{1}{1+y^2}
  +
  \bigg[1-\frac{2}{\pi} \arctan (p N) \bigg]\,
  \delta(y-pN) \quad (y\ge0)
\end{equation}
where $y=pN|\Delta|$. The corresponding cumulative distribution
function is given by
\begin{equation}
  \label{eq:cdf_lorentz}
  \mathrm{cdf}(|\Delta|) = \frac{2}{\pi} \arctan(pN|\Delta|) \quad (0\le|\Delta|\le1)
\end{equation}
and the mean value of $|\Delta|$ is
\begin{equation}
  \label{eq:Delta_av_lorentz}
  \langle |\Delta| \rangle = \frac{1}{\pi} \,
  \frac{1}{p N}
  \ln \big[
  1+(pN)^2
  \big]
  +1
  -\frac{2}{\pi}
  \arctan(p N)
\end{equation}
which has the following limits:
\begin{subequations}
\begin{equation}
  \label{eq:Delta_av_lorentz_small}
    \langle |\Delta| \rangle \simeq 1 - \frac{1}{\pi}\, p N\,,
    \quad
    pN\to0
\end{equation}
and
\begin{equation}
  \label{eq:Delta_av_lorentz_large}
      \langle |\Delta| \rangle \simeq \frac{2}{\pi}\, \frac{\ln (p N)}{p N}\,,
    \quad 
    pN\to\infty\,.
\end{equation}
\end{subequations}
From Fig.~\ref{fig:sw_Delta_mean}, we see that
Eq.\eqref{eq:Delta_av_lorentz} gives a remarkably accurate description
of the empirical $\langle\!\langle|\Delta|\rangle\!\rangle$ vs.~$pN$ behavior over a
large range of $N$ and $p$ values. 

For a comparison of the expression of Eq.~\eqref{eq:cdf_lorentz} for the
cumulative distribution function, we consider only $|\Delta|$-values for
fixed realizations of the random network. Then,
Eq.~\eqref{eq:cdf_lorentz} only appears to be a reasonably accurate
approximation of the numerical results if $p$ is not too small (see
Fig.~\ref{fig:sw_Delta_cdf}). Additional numerical simulations for
$N=100$, $1000$ and $2000$, indeed show a nice collapse of the data on
the curve described by Eq.~\eqref{eq:cdf_lorentz} if $pN$ is larger than
about $100$. A value of $pN \ge 100$ implies that the mean
vertex-to-vertex distance is smaller than about $1/10$ of the mean
distance along the ring, see e.g.\ Fig.~5 of Ref.~\cite{Newman1999a}.
According to the discussion in Ref.~\cite{Newman1999a}, this is the
small-world regime, which suggests that the Lorentz distribution of
Eq.~\eqref{eq:cdf_lorentz} is a good approximation of the
$|\Delta|$-statistics only if the network exhibits small-world behavior.
For very small values of $pN$, we reach the limiting case of only a few
long-range connections and the result, thus, depends on the specific
location of these connections. For larger values of $pN$, however, the
$|\Delta|$-distribution becomes independent of the specific realization
of the network and the system shows self-averaging behavior. Our
numerical results indicate that this self averaging is connected to the
small-world regime.

%For smaller values of $pN$, the results appear to approach the average
%distribution of $|\Delta|$-values in the limiting case of a single
%long-range connection. This distribution can be determined
%analytically from the frequencies of $\Delta(d)$-values shown in
%Table~\ref{tab:ring_lrc_freq}, and it turns out to depend explicitly
%on $N$ and not only on $pN$. In Fig.~\ref{fig:sw_Delta_cdf}, we show
%the limiting average cdf of networks of size $N = 10000$ with a single
%long-range connection. % PROBLEM: this is for the ensemble average
On the other hand, for very large $pN$-values, of the order of $N^2$, we expect that the
cdf-behavior will approach the degenerate one of a fully connected
ring (see Subsection~\ref{sec:toconn}). Preliminary simulations
confirm this trend, but they are very time-consuming and we cannot yet
give a quantitative description of the transition from the
Lorentz-behavior to the behavior of a fully connected ring.

Finally, we note that Eq.~\eqref{eq:Delta_av_lorentz_large} not only
gives a very accurate description of the observed
$\langle\!\langle|\Delta|\rangle\!\rangle$ vs.\ $pN$ behavior
(Fig.~\ref{fig:sw_Delta_mean}), but can also be used to describe the
$pN$-dependence of the
$\langle\!\langle\Delta\!\rangle\!\rangle$-plateaus in
Fig.~\ref{fig:sw_Delta_mean_rd_scaled}. As
Eq.~\eqref{eq:Delta_av_lorentz_large} refers to the mean
value~$\langle|\Delta|\rangle$, it naturally overestimates the plateau
values (see Table~\ref{table:sw_plateau}), but it can also be seen that
the overestimation decreases with increasing~$pN$. This reflects the
fact that with increasing~$pN$, the width of the plateau region
(compared to the region of exponential decay) becomes more and more
dominant.  We also note, however, that the Lorentz-approximation gives
no information on the (exponential) behavior of
$\langle\!\langle|\Delta|\rangle\!\rangle$ vs.~$pd$ for small
$pd$-values.

\section{Summary and Conclusions}

\label{sec:conclusions}

In this paper, we have studied the current redistribution in resistor
networks after the failure of a current-carrying bond. The
corresponding current-redistribution factors~$\Delta_{ij,mn}$
characterize in a convenient way the vulnerability of a flow-carrying
network with respect to cascading failures~\cite{Lehmann2010}.

Information on a network's vulnerability, in the sense of identifying
the most vulnerable links (links with largest voltage drops) can also
be obtained from relations between two nodes, e.g., from the two-point
resistances $R_{ij}$~\cite{Korniss2006}. The current-redistribution
factors $\Delta_{ij,mn}$, however, are relations between two edges and
thus represent a more natural and more adequate tool to study the
vulnerability of a network with respect to flow-induced failure
cascades~\cite{Schaub2014}.

An important property of the current-redistribution
factors~$\Delta_{ij,mn}$ is their dependence on the distance $d$ between
link $i$--$j$ from the failed link $m$--$n$. In this paper, we have thus
determined this dependence $\Delta(d)$ for different types of networks,
and we have studied the influence of network topology (regular networks
vs.~small-world networks) and dimensionality on $\Delta(d)$. In
addition, we have also analyzed the overall statistics of the
$\Delta$-values in the different network topologies. This aspect of the
present investigations was, in particular, motivated by our recent
studies of stochastic load-redistribution
models~\cite{Lehmann2010,Lehmann2010a}.

In Sec.~\ref{sec:crdf}, we have outlined the calculation of the
current redistribution factors $\Delta_{ij,mn}$ in a resistor network
and discussed the use of an interesting relation between
$\Delta_{ij,mn}$ and the two-point resistances $R_{in}$, $R_{im}$,
$R_{jn}$, $R_{jm}$, and $R_{mn}$, see Eq.~\eqref{eq:Delta_from_R}. We note
that this relation is non-linear (because of $R_{mn}$ in the denominator),
i.e., average values of $\Delta_{ij,mn}$ cannot be expressed in terms of average
two-point resistances.

In Sec.~\ref{sec:lattice}, we have calculated the $\Delta$-values for
resistor networks that have the topology of regular lattices
(one-dimensional chain, two-dimensional square lattice, and
three-dimensional simple cubic lattice). As the one-dimensional chain
is a degenerate case, i.e., the $\Delta$-values depend on the boundary
conditions or are even not properly defined, we have also considered
quasi-one-dimensional networks (one-dimensional ladders), for which
these problems do not appear, see Sec.~\ref{sec:quasi_1d}.

For all regular networks, we have derived exact analytic expressions
for $\Delta(d)$, where $d$ denotes the distance of the considered link
from the failed link. It turns out that the behavior of $\Delta(d)$ in
a quasi-one-dimensional ladder is completely different from that in
higher-dimensional regular networks.

For quasi-one-dimensional lattices, $|\Delta|$ decays exponentially with
increasing distance d from the failed link, $|\Delta|(d) \simeq
\exp(-\chi d)$, whereas for higher dimensional lattices we observe a
power law decay, $|\Delta|(d) \simeq 1/d^2$ for the two-dimensional
square lattice and $|\Delta|(d) \simeq 1/d^3$ for the three-dimensional
simple cubic lattice. Apart from the overall power law decay, the
$|\Delta|$-behavior in regular two- and three-dimensional lattices
exhibits a characteristic angular dependence. We further observe that
the dominating behavior of the overall distribution function,
$\varrho(|\Delta|)$, is the same for all regular lattices of dimension
two and higher, $\varrho(|\Delta|) \simeq 1 / |\Delta|^2$. We finally
remark that breakdown models with a similar power-law dependence of the
stress redistribution have been studied in Refs.~\cite{Hidalgo2002} and
\cite{Lehmann2010a}. Interestingly enough, if we compare the
$|\Delta|^{-2}$ load redistribution with the one of the stochastic
fibre-bundle model studied in Ref.~\cite{Lehmann2010a}, we find that it
would correspond to a range-dependent load sharing with a
$r^{-2}$-dependence, which is just at the border of the transition
between global and local load sharing~\cite{Hidalgo2002}.

In random small-world networks the behavior of the current
redistribution factors $\Delta_{ij,mn}$ becomes much more complex
(Section~\ref{sec:random_small_world}). We consider networks that are
constructed from a one-dimensional ring by adding random long-range
links with probability $2p/N$ (Newman and Watts, see
Ref.~\cite{Newman1999a}), and we have performed a large number of
numerical simulations for network sizes up to $N = 10’000$ sites and
for $p$-values in the range $0.001 < p <1$.

The behavior of the mean value of $|\Delta|$ as a function of the ring
distance~$d$ turns out to be distinctly different from that observed in
regular networks. For small values of $pd$,
$\langle\!\langle|\Delta|\rangle\!\rangle$
decreases exponentially with $d$, i.e., the behavior is similar to that
of a quasi-one-dimensional network. Around some critical value, $d =
d_\mathrm{c}(p, N)$, however, $\langle\!\langle|\Delta|\rangle\!\rangle$ saturates and
stays constant for larger values of $pd$. It is interesting to note that
the corresponding behavior of $-\ln
\langle\!\langle|\Delta|\rangle\!\rangle$ vs. $d$ is
very similar to that of the graph distance~ $d_\mathrm{graph}$
(shortest-path distance) vs.\ $d$, see e.g. Ref.~\cite{Newman1999a}. In
small-world networks, $|\Delta|$ thus appears to depend exponentially on
the shortest-path distance to the failed link. Note that we have
observed a similar behavior already in the limiting case of a single
long-range connection, but there only if $|\Delta|$ is averaged over all
possible lengths and positions of the long-range link. In random
small-world networks, however, this behavior is independent of the
specific realization, at least if $pN$ is not too small.

Another interesting aspect of the current redistribution in
small-world networks is the overall statistics of the
$|\Delta|$-values. A careful analysis of our numerical results has
revealed that the observed dependences can very accurately be
described by assuming that the statistics of the $pN|\Delta|$-values
is given by a half-Lorentz distribution of width $\gamma = 1$,
Eq.~\eqref{eq:rho_lorentz}. A detailed analysis of the cumulative
distribution function $\mathop\mathrm{cdf}(|\Delta|)$ shows, however,
that it follows a Lorentz distribution only if $pN \gtrsim 100$,
i.e. if the network exhibits small-world behavior, see
Fig.~\ref{fig:sw_Delta_cdf} and the corresponding remarks in the
text. For smaller values of $pN$, the results for
$\mathop\mathrm{cdf}(|\Delta|)$ depend on the specific realization of
the randomly chosen long-range connections. If $pN \gtrsim 100$ (but
not too close to the fully connected limit), this is no longer the
case, i.e., the system shows self-averaging behavior.

These observations are very interesting in several respects. They show
that the transition to a self-averaging behavior occurs at precisely
the $pN$-value, above which the $|\Delta|$-distribution is described by a
(half-)Lorentz distribution. The Lorentz distribution, in turn, has a fat
tail, $\simeq 1/|\Delta|^2$, and this is a very important aspect for the onset
and propagation of failure cascades. The $|\Delta|$-distributions of regular
lattices in two- and three dimensions, in fact, also exhibit a $1 /
|\Delta|^2$ tail. We note, however, that in these cases this behavior of $\varrho(|\Delta|)$ is an
artefact of the $d$-dependence of $|\Delta|$. In small-world networks, on the
other hand, the graph-distance is the same for almost all links that
are not very close to the failed link, so that here the statistics of
$|\Delta|$-values is a more meaningful concept.

Small-world networks are often used as models for real-world
infrastructure networks, e.g., power grids. The $|\Delta|$-statistics of
such networks could thus be an adequate concept to describe the
influence of a failed connection on the other connections, and we have
seen that in small-world networks, this influence is almost independent
of the geometrical location of the affected connections. This gives some
justification to the use of stochastic models, such as proposed in
Ref.~\cite{Lehmann2010}, to describe cascading failure propagation in
real-world networks.

The analysis of such stochastic breakdown models is, however, beyond the
scope of the present paper. Interesting questions in this context would
be the corresponding dependence of breakdown properties on the
load-redistribution statistics. Similar investigations have been
performed for local load-sharing rules in fibre-bundle models, see e.g.\
Refs.~\cite{Sinha2015} and \cite{Kim2005}, where a transition to a
mean-field behavior of the system was observed as a function of the
dimensionality of the system (for regular lattices) or the rewiring
probability (for small-world networks), respectively. It is an open
question whether such transitions can also be observed in resistor
networks, where the current-redistribution, as given by Kirchhoff's
laws, is inherently non-local.

\appendix

\section{Derivation of Eq.~\eqref{eq:Delta_from_R}}

\label{sec:app_delta}

In the following, we prove Eq.\eqref{eq:Delta_from_R} and start by writing the
solution to Eq.\eqref{eq:kirchhoff} in the form
\begin{equation}
  \label{eq:solution_kirchhoff}
        U = \mathsf{X} I^\mathrm{s},
\end{equation}
where $\mathsf{X}$ denotes the (pseudo-)inverse of the nodal
admittance matrix~$\mathsf{Y}$. A removal of bond $m$--$n$ can now be modeled (as far as
the rest of the network is concerned) by adding suitably chosen
current injections $\Delta I^\mathrm{s}_m = - \Delta I^
\mathrm{s}_n$ (see, e.g., Ref.~\cite{Wood1996},
Appendix 11 A). If $\Delta I^\mathrm{s}_m$ is chosen such that
\begin{equation}
  \label{eq:iprime1}
        \Delta I^\mathrm{s}_m = I'_{mn}\,,
\end{equation}
where $I'_{mn}$ is equal to the current flow through bond $m$--$n$, there is no current flowing from nodes $m$ and $n$ to the rest of the network. As far as the rest of the network is concerned, this is the same as a removal of bond $m$--$n$.

With Eq.~\eqref{eq:solution_kirchhoff} we can now express the change
of voltage~$\Delta U_k$ in an arbitrary node~$k$ in terms of the
injected currents~$\Delta I^\mathrm{s}_m= - \Delta I^
\mathrm{s}_n$ as
\begin{equation}
  \label{eq:delta_u_k}
        \Delta U_k = (X_{km} - X_{kn}) \Delta I^\mathrm{s}_m\,.
\end{equation}
Using Ohm's law, this yields, on one hand, the new current through
bond $m$--$n$:
\begin{equation}
\begin{split}
  \label{eq:iprime2}
  I'_{mn} & {}=
  I_{mn} + \frac{\Delta U_m - \Delta U_n}{r_{mn}}
  \\
  & {} = 
  I_{mn} + \frac{X_{mm} + X_{nn} - 2 X_{mn}}{r_{mn}}\,
  \Delta I^\mathrm{s}_m\,.
\end{split}
\end{equation}
Together with condition~\eqref{eq:iprime1}, this gives an expression for
the current injections in terms of the original current~$I_{mn}$ through bond $m$--$n$:
\begin{equation}
  \label{eq:delta_ism}
  \Delta I^\mathrm{s}_m = \frac{1}{1 - (X_{mm} + X_{nn} - 2 X_{mn})/r_{mn}}
    I_{mn}\,.
\end{equation}

On the other hand, we can again use Ohm's law in the
definition~\eqref{eq:delta_def} of the current-redistribution factor
%$\Delta_{ij,mn}$
\begin{equation}
  \Delta_{ij,mn} = \frac{I'_{ij} - I_{ij} }{I_{mn}} =
    \frac{(\Delta U_i - \Delta U_j)/r_{ij}}{I_{mn}}
\end{equation}
and obtain, together with Eqs.~\eqref{eq:delta_u_k} and
\eqref{eq:delta_ism}, the result
\begin{equation}
  \label{eq:delta_2}
  \Delta_{ij,mn} = \frac{(X_{im} - X_{in} - X_{jm} + X_{jn})/r_{ij}}
  {1 -  (X_{mm} + X_{nn} -2 X_{mn})/r_{mn}}\,.
\end{equation}
Finally, with Eqs.~\eqref{eq:two_point_R},
\eqref{eq:current_injections}, and \eqref{eq:solution_kirchhoff}, we find
\begin{equation}
	\label{eq:R_from_X}
        R_{ij} = X_{ii} + X_{jj} - 2 X_{ij}\,,
\end{equation}
and Eq.\eqref{eq:delta_2} can thus be written as
\begin{equation}
 \Delta_{ij,mn} = \frac{1}{2 r_{ij}}
 \frac{R_{in} - R_{im} +  R_{jm} - R_{jn}}{1 -  R_{mn}/r_{mn}}\,.
\end{equation}

\section{Exact values of $\Delta_\parallel(x,y)$ and $\Delta_\perp(x,y)$ for an Infinite 2d Square Lattice}

\label{sec:app_2d_square}

Using Eq.~\eqref{eq:Delta_from_R}, we can calculate exact
$\Delta(x,y)$-values from exact values for the two-point resistances~$R(x,y)$. For an infinite square lattice, a table of such values [for $0\le x, y \le 20$] has been compiled by S.\ and R.~Hollos~\cite{Hollos2005_Ref3}. In Table~\ref{tab:2d_Delta}, we list a few examples of exact expressions for $|\Delta_\parallel|(x,y)|$ and $|\Delta_\perp|(x,y)$.
\begin{table}[b]
\renewcommand{\arraystretch}{2.5} \centering
\begin{tabular}{l@{\hspace{1cm}}l}
$\displaystyle|\Delta_\parallel|(0,1) = \frac{4}{\pi} - 1$         & $\displaystyle|\Delta_\perp|(0,0) = 1-\frac{2}{\pi}$\\
$\displaystyle|\Delta_\parallel|(0,2) = \frac{16}{\pi} - 5$        & $\displaystyle|\Delta_\perp|(0,1) = 2-\frac{6}{\pi}$\\
$\displaystyle|\Delta_\parallel|(0,3) = \frac{236}{3\pi} - 25$     & $\displaystyle|\Delta_\perp|(0,2) = 10-\frac{94}{3\pi}$\\
$\displaystyle|\Delta_\parallel|(0,4) = \frac{1216}{3\pi} - 129$   & $\displaystyle|\Delta_\perp|(0,3) = 52-\frac{490}{3\pi}$\\
$\displaystyle|\Delta_\parallel|(0,5) = \frac{32092}{15\pi} - 681$ & $\displaystyle|\Delta_\perp|(0,4) = 276-\frac{13006}{15\pi}$\\
$\displaystyle|\Delta_\parallel|(1,0) = \frac{4}{\pi} -1 $ & $\displaystyle|\Delta_\perp|(1,0) = 1-\frac{2}{\pi}$\\
$\displaystyle|\Delta_\parallel|(1,1) = 0$                 & $\displaystyle|\Delta_\perp|(1,1) = 2-\frac{6}{\pi}$\\
$\displaystyle|\Delta_\parallel|(1,2) = 3- \frac{28}{3\pi}$
\end{tabular}
\caption{Exact expressions for $|\Delta_\parallel|(x,y)$ and
$|\Delta_\perp|(x,y)$ in an infinite square lattice.}
\label{tab:2d_Delta}
\end{table}

\end{document}